\documentclass{ifacconf}
\ifdefined\pdfminorversion\pdfminorversion=4\fi
\usepackage[T1]{fontenc}
\usepackage[utf8]{inputenc}
\usepackage{lmodern}
\usepackage{amsmath,amssymb,mathtools}
\usepackage{graphicx}
\usepackage{url}
\usepackage{natbib}
\usepackage{booktabs}
\usepackage{microtype}
\usepackage{tikz}
\usetikzlibrary{arrows.meta,positioning,calc}

\newcommand{\R}{\mathbb{R}}

\begin{document}
\begin{frontmatter}
\title{Comparing model representations and fitting methods for port-Hamiltonian system identification\thanksref{funding}}
\thanks[funding]{This work was supported by the Research Council of Norway through the project PhysML: Structure-based machine learning for physical systems (project no.\ 338779).}
\author[SINTEF]{Sølve Eidnes}
\author[NTNU]{Håkon Noren Myhr}
\address[SINTEF]{Department of Mathematics and Cybernetics, SINTEF Digital,
Oslo, Norway (e-mail: solve.eidnes@sintef.no).}
\address[NTNU]{Department of Mathematical Sciences, Norwegian University of Science and Technology,
Trondheim, Norway.}
\begin{abstract}
Port-Hamiltonian models describe physical systems by separating stored energy, dissipation and external inputs. Learning such models from data requires a choice of representation and fitting method. We compare using neural networks, analytic expressions and combinations of the two for representing the different components of the port-Hamiltonian system, and compare fitting the models with gradient-based optimisation or sparse regression. Testing on simulations of mechanical and electrical systems, with and without added noise, we present prediction accuracy and computational cost of the different approaches.
\end{abstract}
\end{frontmatter}

\section{Introduction}
Port-Hamiltonian models describe dynamics using separate functions for stored energy, dissipation and coupling to the surroundings through external inputs \citep{vanderschaft}. To identify systems of this form from data one must choose how to represent the unknown functions and how to fit them to observations. Preserving the port-Hamiltonian structure allows the learned energy, dissipation and input map to be interpreted separately and provides a basis for control.

The unknown functions can be represented by analytic expressions obtained by fitting combinations of prescribed candidate functions \citep{lee,holmsen,myhr}. Neural-network representations have been considered for the same purpose \citep{desai,eidnes}, building on Hamiltonian neural networks \citep{greydanus}. Hybrid representations combine analytic expressions for some functions with neural networks for others \citep{holmsen,cherifi,khosrovian}.

Sparse regression is used to select a small set of terms in a differential equation from a candidate library \citep{brunton} and has been adapted to jointly identify the energy and dissipation of port-Hamiltonian systems \citep{myhr}. Gradient-based optimisation is widely used to fit neural networks but has also been used to fit the coefficients of analytic port-Hamiltonian models \citep{lee,holmsen}. We compare two variants of each of these approaches, giving four different fitting methods. Our gradient-based optimisation minimises either integration-scheme residuals or errors in simulated trajectories. Our sparse-regression fits alternate ridge-regularised linear least-squares solves with term removal, using estimated derivatives or a midpoint discretisation.

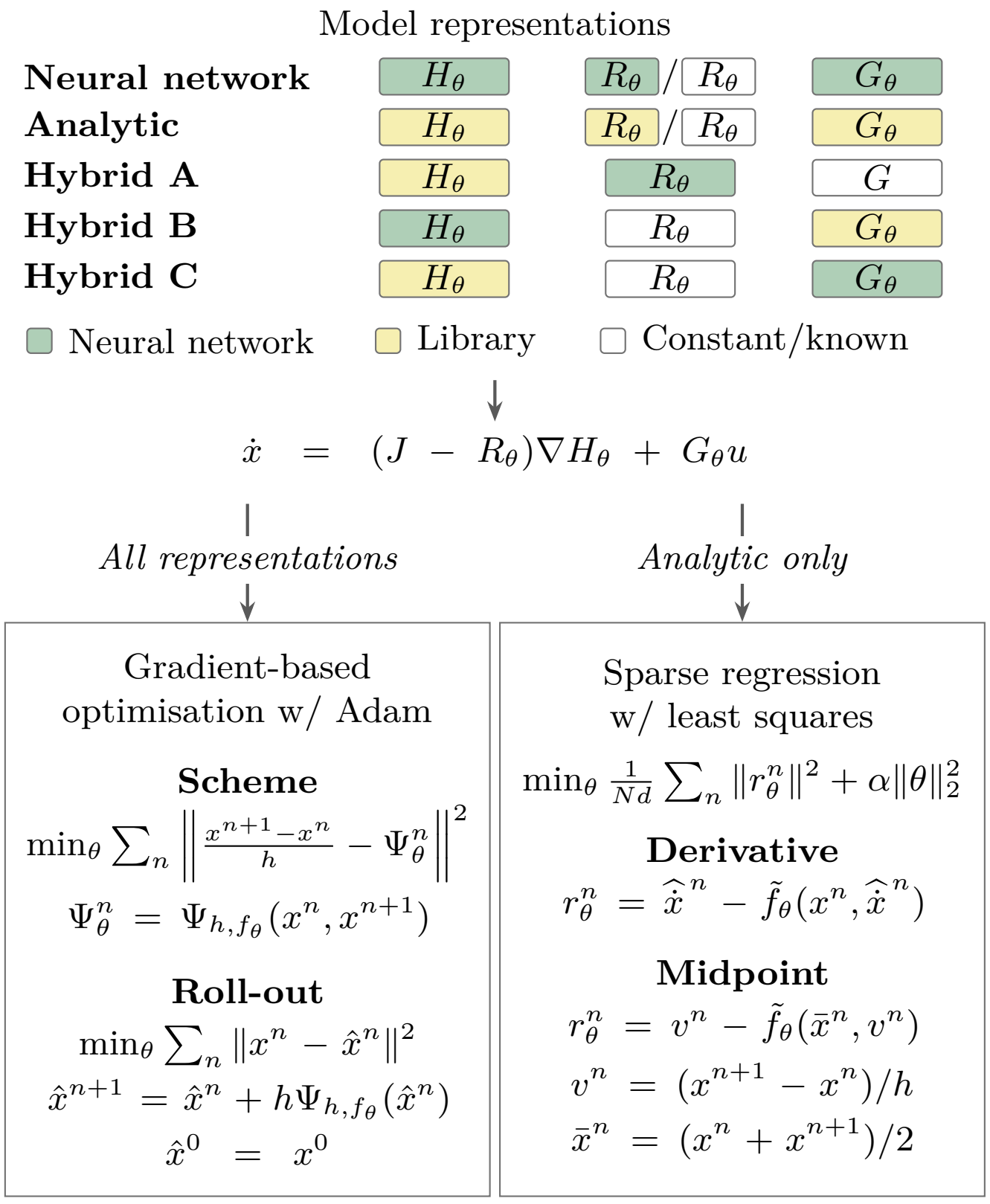
\begin{figure}[t]
\centering
\definecolor{overviewNetwork}{HTML}{AFCFB7}
\definecolor{overviewLibrary}{HTML}{F6EFB1}
\begin{tikzpicture}[
  x=1cm,y=1cm,
  every node/.style={font=\normalfont\fontsize{9}{10}\selectfont,align=center,inner sep=2pt},
  component/.style={draw=black!55,rounded corners=1pt,minimum width=1.15cm,minimum height=0.32cm,inner sep=1pt},
  network/.style={component,fill=overviewNetwork},
  library/.style={component,fill=overviewLibrary},
  fixed/.style={component,fill=white},
  choice/.style={minimum width=0.65cm},
  method/.style={draw=black!55,text width=4.1cm,minimum height=5.1cm,inner sep=3pt,anchor=north},
  arrow/.style={-{Stealth[length=1.8mm]},black!65,line width=0.8pt}
]
\node[inner sep=0pt] at (0,3.1) {Model representations};
\node[anchor=west,inner sep=0pt] at (-4.2,2.65) {\textbf{Neural network}};
\node[anchor=west,inner sep=0pt] at (-4.2,2.2) {\textbf{Analytic}};
\node[anchor=west,inner sep=0pt] at (-4.2,1.75) {\textbf{Hybrid A}};
\node[anchor=west,inner sep=0pt] at (-4.2,1.3) {\textbf{Hybrid B}};
\node[anchor=west,inner sep=0pt] at (-4.2,0.85) {\textbf{Hybrid C}};

\node[network] at (-0.45,2.65) {$H_\theta$};
\node[network,choice] at (1.13,2.65) {$R_\theta$};
\node at (1.56,2.65) {/};
\node[fixed,choice] at (1.99,2.65) {$R_\theta$};
\node[network] at (3.4,2.65) {$G_\theta$};
\node[library] at (-0.45,2.2) {$H_\theta$};
\node[library,choice] at (1.13,2.2) {$R_\theta$};
\node at (1.56,2.2) {/};
\node[fixed,choice] at (1.99,2.2) {$R_\theta$};
\node[library] at (3.4,2.2) {$G_\theta$};
\node[library] at (-0.45,1.75) {$H_\theta$};
\node[network] at (1.56,1.75) {$R_\theta$};
\node[fixed] at (3.4,1.75) {$G$};
\node[network] at (-0.45,1.3) {$H_\theta$};
\node[fixed] at (1.56,1.3) {$R_\theta$};
\node[library] at (3.4,1.3) {$G_\theta$};
\node[library] at (-0.45,0.85) {$H_\theta$};
\node[fixed] at (1.56,0.85) {$R_\theta$};
\node[network] at (3.4,0.85) {$G_\theta$};

\node[network,minimum width=0.22cm,minimum height=0.22cm,inner sep=0pt] at (-4.05,0.3) {};
\node[anchor=west,inner sep=0pt] at (-3.8,0.3) {Neural network};
\node[library,minimum width=0.22cm,minimum height=0.22cm,inner sep=0pt] at (-0.95,0.3) {};
\node[anchor=west,inner sep=0pt] at (-0.7,0.3) {Library};
\node[fixed,minimum width=0.22cm,minimum height=0.22cm,inner sep=0pt] at (1.05,0.3) {};
\node[anchor=west,inner sep=0pt] at (1.3,0.3) {Constant/known};

\node[text width=8.25cm,minimum height=0.55cm] (model) at (0,-0.7)
 {$\dot x=(J-R_\theta)\nabla H_\theta+G_\theta u$};
\draw[arrow] (0,-0.05) -- (model.north);

\node[method] (gradient) at (-2.2,-2.2)
 {Gradient-based\\optimisation w/ Adam\\[7pt]
  {\fontsize{9}{10}\selectfont
   \textbf{Scheme}\\[2pt]
   $\min_\theta\sum_n\left\|\frac{x^{n+1}-x^n}{h}-\Psi_\theta^n\right\|^2$\\[3pt]
   $\Psi_\theta^n=\Psi_{h,f_\theta}(x^n,x^{n+1})$\\[8pt]
   \textbf{Roll-out}\\[2pt]
   $\min_\theta\sum_n\|x^n-\hat x^n\|^2$\\[3pt]
   $\hat x^{n+1}=\hat x^n + h\Psi_{h,f_\theta}(\hat x^n)$\\[3pt]
   $\hat x^0=x^0$}};
\node[method] (regression) at (2.2,-2.2)
 {Sparse regression\\w/ least squares\\[5pt]
  {\fontsize{9}{10}\selectfont
   $\min_\theta\frac{1}{Nd}\sum_n\|r_\theta^n\|^2+\alpha\|\theta\|_2^2$\\[7pt]
   \textbf{Derivative}\\[2pt]
   $r_\theta^n=\widehat{\dot x}^{\,n}-\tilde f_\theta(x^n,\widehat{\dot x}^{\,n})$\\[7pt]
   \textbf{Midpoint}\\[2pt]
   $r_\theta^n=v^n-\tilde f_\theta(\bar x^n,v^n)$\\[4pt]
   $\displaystyle v^n=(x^{n+1}-x^n)/h$\\[3pt]
   $\displaystyle\bar x^n=(x^n+x^{n+1})/2$}};
\draw[arrow] (-2.2,-1.15) -- (gradient.north);
\draw[arrow] (2.2,-1.15) -- (regression.north);
\node[fill=white,inner xsep=3pt,inner ysep=2pt] at (-2.2,-1.65) {\textit{All representations}};
\node[fill=white,inner xsep=3pt,inner ysep=2pt] at (2.2,-1.65) {\textit{Analytic only}};
\end{tikzpicture}
\caption{Overview of the model representations and fitting objectives compared in the paper. Here $x^n$ is an observed state at time $t^n$, $\widehat{\dot x}^{\,n}$ its estimated derivative, and $h=t^{n+1}-t^n$ the sampling interval. For the model field $f_\theta = (J-R_\theta)\nabla H_\theta+G_\theta u$, a numerical integration scheme approximates the state change over one step by $h\Psi_{h,f_\theta}$. Regression uses $\tilde f_\theta(x,v)$: the analytic model $f_\theta$, with $\nabla_p H_\theta$ in the damping term replaced by the corresponding components of the observed rate $v$. For regression, $N$ is the number of samples, $d$ the number of state variables, and $\alpha\geq0$ controls the strength of the ridge penalty.}
\label{fig:comparison}
\end{figure}

In this paper, we compare the three types of representation and the two families of fitting methods, summarised in Figure~\ref{fig:comparison}, to examine how the choices affect prediction accuracy and computational cost when applied to data from simulated mechanical and electrical systems. We briefly consider how well the different models have learned the input gain, but a broader assessment of the identified functions is outside the scope of this study. This can, however, be explored using our public code on \url{https://github.com/SINTEF/PHSI}.

\section{Models and fitting methods}
\subsection{Port-Hamiltonian formulation}
Let \(x(t)\in\R^d\) be the state, \(u(t)\) a scalar input and \(H(x)\) the Hamiltonian. With a skew-symmetric interconnection matrix \(J\), a symmetric positive semidefinite dissipation matrix \(R(x)\) and an input map \(G(x)\), the port-Hamiltonian formulation is
\begin{equation}
 \begin{aligned}
 \dot x&=(J-R(x))\nabla H(x)+G(x)u,\\
 y&=G(x)^\top\nabla H(x),
 \end{aligned}
 \label{eq:ph}
\end{equation}
where \(y\) is the output conjugate to \(u\), and \(uy\) is the supplied power \citep{vanderschaft}. Since \(J^\top=-J\), the energy balance is \(\dot H=-\nabla H^\top R\nabla H+uy\). If \(H\) is bounded from below, a constant shift gives a nonnegative storage function and the system is passive.

Given measurements of the state and the input, port-Hamiltonian system identification means identifying representations \(H_\theta\), \(R_\theta\) and, when unknown, \(G_\theta\), where \(\theta\) represents the learnable parameters. In this paper, we keep \(J\) fixed and constrain \(R_\theta\) to be diagonal and positive semidefinite. For an unknown input map, \(G_\theta(x)=e\,g_\theta(z(x))\), where the standard unit vector \(e\) selects the known actuated coordinate and \(z(x)\) selects the prescribed state variable on which the gain depends. Only the gain \(g_\theta\) is learned. The learned input map \(G_\theta\) is used in both the state equation and the output equation in \eqref{eq:ph}. 

\subsection{Model representations}\label{sec:representations}
In the \emph{neural-network representation}, \(H_\theta\) is a neural network, as in \citet{desai,eidnes,duong}. We use two hidden layers of 64 hyperbolic tangent units and a linear output. For constant damping, the diagonal coefficients on the damped coordinates are learned as nonnegative parameters in \(\theta\), while the other entries of \(R_\theta\) are fixed at zero. For state-dependent damping, each damped coordinate has a separate network taking only that coordinate as input, with two hidden layers of 16 hyperbolic tangent units and a softplus output, ensuring a positive coefficient. An unknown input gain is represented by a network with two hidden layers of 16 hyperbolic tangent units and a linear output, and takes only the selected state variable as input. 

In the \emph{analytic representation}, \(H_\theta(x)=\sum_k a_k\varphi_k(x)\), where the candidate functions \(\varphi_k\) are predefined and the coefficients \(a_k\) are learned. This candidate-library approach, popularised by \citet{brunton}, was combined with port-Hamiltonian formulations in \citet{lee,myhr}. We test different libraries for the candidate functions. Each state-dependent damping coefficient is represented by a nonnegative combination of nonnegative candidate functions, while the coefficients for the Hamiltonian and input gain are unconstrained. The libraries in our main experiments are specified in sections~3.1 and~3.2, while 
alternative libraries are tested in Section~\ref{sec:libraries}.

The \emph{hybrid representations} combine analytic and neural-network terms, as also considered by \citet{holmsen,cherifi,khosrovian}. For state-dependent damping with a known input map, \emph{hybrid A} combines an analytic Hamiltonian model with the above-mentioned neural networks for the damping. For constant damping but an unknown input map, \emph{hybrid B} combines a neural network model of the Hamiltonian with an analytic model of the input gain, while \emph{hybrid C} combines an analytic Hamiltonian model with a neural-network input gain. Both learn the constant damping coefficients.

\subsection{Fitting methods}\label{sec:fitting}
We compare two families of fitting methods, each with two variants. Gradient-based optimisation is used for every representation, while sparse regression is used only for the fully analytic representation.

\subsubsection{Gradient-based optimisation}
We use the adaptive moment estimation (Adam) optimiser to minimise a mean squared loss. Our two variants of the loss differ in how they compare the model output with the observations. \textit{Integration-scheme fitting} measures how closely pairs of adjacent observed states satisfy a symmetric fourth-order integration scheme \citep{eidnes,holmsen}. The scheme is mono-implicit, meaning it is explicitly given by the two end points, so evaluating the residual requires neither trajectory simulation nor a separate estimate of the state derivatives \citep{celledoni}. \textit{Roll-out fitting} instead simulates the model over short windows, starting from the first observed state in each window, and compares its predictions with subsequent observations of the state variables. We use explicit fourth-order Runge--Kutta integration and differentiate through its steps to obtain the parameter gradients \citep{chen,duong,moradi}; our implementation builds on that of the neural Hamiltonian ordinary differential equations of \citet{meltzer}.

\subsubsection{Sparse regression}
Sparse regression fits the analytic coefficients by linear least squares while constraining the damping coefficients to be nonnegative \citep{brunton}. Following \citet{myhr}, we include a ridge penalty on the squared coefficients to limit large coefficient values. For our canonical systems, we write \(x=(q,p)\), with positions or charge \(q\), momenta or flux \(p\), and damping and input acting only on \(p\). The relation \(\dot q=\nabla_p H\) then lets us replace this energy derivative in the damping term by a rate estimated from the observations. This avoids products of unknown energy and damping coefficients and makes the joint fit linear \citep{myhr}. In the \textit{derivative fit} variant, the derivatives of the states are estimated using second-order central differences for raw data and local-polynomial differentiation for smoothed data. The \textit{midpoint fit} variant instead evaluates the candidate functions halfway between consecutive observed states, using the change between these states divided by the sampling interval as the corresponding rate. This formulation is based on the implicit midpoint integration scheme. To obtain a sparse model, we alternate these penalised fits with the removal of terms with small coefficients, as described in Section~\ref{sec:pruning}.

\section{Numerical experiments}\label{sec:examples}
\subsection{Two-mass system}
Consider two masses connected by a nonlinear spring, with another nonlinear spring connecting the first mass to a fixed support. Let \(q_1,q_2\) be their displacements from equilibrium, \(p_1,p_2\) their momenta and \(m_1,m_2\) their masses. With spring parameters \(k_0,k_1,b_0,b_1\), the Hamiltonian of the system is \(H=p_1^2/(2m_1)+p_2^2/(2m_2)+[k_0q_1^2+k_1(q_2-q_1)^2]/2+[b_0q_1^4+b_1(q_2-q_1)^4]/4\). An external force \(g_2(q_2)u\) acts on the second mass, where \(u\) is the measured input and \(g_2\) is the input gain. With damping coefficients \(d_i\) and input gains \(g_i\), the equations of motion are
\begin{equation}
 \begin{aligned}
 \dot q_i&=p_i/m_i,\\
 \dot p_i&=-\frac{\partial H}{\partial q_i}-d_i(p_i)p_i/m_i+g_i(q_2)u,
 \end{aligned}
\end{equation}
for \(i=1,2\), with \(g_1=0\). We use \((m_1,m_2)=(1,1.2)\), \((k_0,k_1)=(1,1.5)\) and \((b_0,b_1)=(0.5,0.3)\) in our experiments. For both mechanical examples, the analytic Hamiltonian library contains all nonconstant monomials in \(q_1,q_2,p_1,p_2\) of total degree at most four.

In the first example, we learn the Hamiltonian, two constant damping coefficients and an input gain that depends only on the second mass's position. The ground truth has constant damping, \(d_1=0.15\) and \(d_2=0.20\), and gain \(g_2(q_2)=1+0.5\tanh(q_2/\ell)\), with displacement scale \(\ell=0.5\). The library for the analytic model of the gain contains a constant and powers of \(q_2\) through degree four and cannot represent the true gain exactly.

In the second example we assume the constant gain \(g_2=1\) is known, but we have to learn the Hamiltonian and state-dependent damping, whose ground truth is \(d_i(p_i)=c_i+a_i\exp[-p_i^2/s_i^2]\), with \(c=(0.15,0.20)\), \(a=(0.25,0.20)\) and \((s_1,s_2)=(0.5,1)\). For the analytic model of the damping function, we use a library containing a constant, \(p_i^2\) and Gaussian functions \(\exp[-(p_i-\mu)^2/s^2]\). The centres \(\mu\) are seven equally spaced points in \([-1.5,1.5]\), and the width scale \(s\) takes the values \(0.3\) and \(0.6\). This gives 16 candidate functions per damping coefficient, but does not include the true scales.

\subsection{Nonlinear electrical circuit}
The electrical example is a series circuit with an unknown, state-dependent input gain. Let \(q\) be capacitor charge, \(\phi\) inductor flux, \(C>0\) capacitance, \(r>0\) resistance and \(I_s,\delta>0\) current and flux scales. The Hamiltonian is \(H(q,\phi)=q^2/(2C)+I_s\delta\log\cosh(\phi/\delta)\). With measured voltage \(u\) and charge-dependent gain \(g(q)\), the governing equations are
\begin{equation}
 \begin{aligned}
 \dot q&=I_s\tanh(\phi/\delta),\\
 \dot\phi&=-q/C-rI_s\tanh(\phi/\delta)+g(q)u.
 \end{aligned}
\end{equation}
We use the nonlinear-inductor law of \citet{castanos} in a simpler two-state circuit, with \(C=I_s=1\), \(r=0.15\) and \(\delta=0.7\). The ground truth input gain is \(g(q)=0.5+[1+(q/q_0)^2]^{-1}\), with charge scale \(q_0=0.75\). Writing the inductor current as \(i=I_s\tanh(\phi/\delta)\), the conjugate output is \(y=g(q)i\), and \(\dot H=-ri^2+uy\). Our models learn the Hamiltonian, the constant resistance and the state-dependent input gain. The analytic Hamiltonian library contains all nonconstant monomials in \(q,\phi\) of total degree at most four, and the gain library contains \(1,q,q^2,q^3,q^4\). Neither library can represent its corresponding reference function exactly.

\subsection{Data}
For each example, we use three independent datasets, each containing 24 training trajectories, eight validation trajectories and 16 test trajectories. Training and validation trajectories go from $t=0$ to $t=20$ with sampling interval $h=0.05$, while the test trajectories extend to \(t=40\). All three examples use smooth, time-varying input signals.

For the examples with unknown input maps, we simulate two trajectories from each initial state, using inputs of opposite sign. This helps distinguish the effect of the input from the internal dynamics. All fitting methods use the same datasets, with separate trajectories for training, validation and testing.

Table~\ref{tab:three-noise} compares five conditions: noise-free data, and data with added noise of two different levels, each with and without smoothing. For the noisy conditions, we add independent, zero-mean Gaussian noise to the training and validation state observations. For each state variable, the noise standard deviation is 1\% or 5\% of its noise-free training standard deviation. The applied input and the test reference trajectories are noise-free.

\subsection{Fitting procedure}
\label{sec:pruning}
We fit each model to the training data and use validation data to select among candidate fits. For models with analytic functions, the procedure includes \emph{pruning}: terms with small coefficients, scaled by their contribution to the state equations, are removed and the remaining parameters are adjusted. In steps, the fitting procedure is:
\begin{enumerate}
\item \emph{Prepare the data.} Use raw training data or apply an 11-sample cubic local-polynomial smoother, discarding the first and last five samples of each smoothed trajectory. In both cases, determine normalisation from the raw training data.
\item \emph{Fit the full model.} Fit the parameters using the chosen method from Section~2.3. For regression, we test ridge strengths \(\alpha\in\{0,10^{-8},10^{-6},10^{-5},10^{-4},10^{-2}\}\), where \(\alpha=0\) means no regularisation.
\item Only if the model contains analytic library terms, \emph{prune and adjust the remaining parameters.} Remove terms whose scaled coefficients have absolute values below a chosen threshold. For both fitting methods, test thresholds \(0,10^{-4},10^{-3},10^{-2},10^{-1}\), where zero means no pruning. For models with both an analytic Hamiltonian and an analytic damping or input-gain library, test all combinations of the two thresholds.
\begin{enumerate}
\item Gradient-based optimisation: For each threshold combination, start from the same full-library model selected by validation, retain the fitted values of the remaining parameters and reset only the optimiser state before continuing optimisation.
\item Sparse regression: For each combination of ridge strength and pruning thresholds, start with a fit using all library terms. Remove terms below the thresholds and fit the remaining coefficients again, retaining the same ridge penalty. Repeat until no further terms are removed, with a limit of 20 iterations.
\end{enumerate}
\item \emph{Select and evaluate.} Choose the candidate with the lowest prediction error on the validation data. For regression, this jointly selects the ridge strength and pruning thresholds. Evaluate the selected model on the test trajectories.
\end{enumerate}

For gradient-based optimisation, the initial fits have time limits of 30 minutes for scheme fitting and 90 minutes for roll-out fitting. For models with analytic library terms, each pruning candidate, including the unpruned model, receives a further five or fifteen minutes, respectively, for continued optimisation. Regression has a single 50-minute limit per dataset for the complete search over ridge strengths and pruning thresholds.

\section{Results and discussion}

\subsection{Comparison of representations and fitting methods}

Our main experiment is reported in Table~\ref{tab:three-noise}, which compares prediction errors and fitting times for the three examples, using the candidate-function libraries described in Section~\ref{sec:examples}. We evaluate predictions on the test trajectories up to \(t=20\) using the normalised root-mean-square error (NRMSE). The normalisation is done by dividing by the standard deviation in the noise-free training data for each state variable. The reported errors are the arithmetic mean NRMSE over the three datasets. The fitting times reported in the table include preprocessing, validation, initial full-library fitting and the complete candidate search, but exclude inference on the test data. These times are averaged over all five conditions and three datasets, including unsuccessful fits. Each fit used two processor threads on an Intel Core i9-14900KF.

\begin{figure}[tb]
\centering
\includegraphics{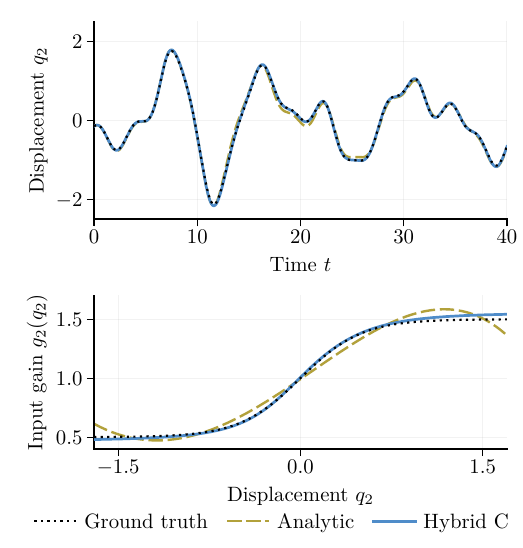}
\caption{Ground truth and model predictions for the first test trajectory (top) and input gain (bottom), for the first dataset on the two-mass system with constant damping. The predictions are from the models fitted to smoothed data with 5\% noise.}
\label{fig:learned-input-gain}
\end{figure}

Figure~\ref{fig:learned-input-gain} shows the predicted displacement of the second mass and the learned input gain function in the two-mass system with constant damping and unknown input gain, from a purely analytic model and a hybrid model with an analytic model of the Hamiltonian and a neural-network input gain model. 
Both models are fitted by gradient-based optimisation using the integration-scheme loss. Although the hybrid model gives a lower prediction error, its advantage is perhaps most apparent in how accurately it learns the gain function.

\begin{table*}[t]
\centering
\caption{Prediction error on \(t\in[0,20]\) and mean fitting time for the different methods.}
\label{tab:three-noise}
\setlength{\tabcolsep}{4pt}
\begin{tabular}{llrrrrrr}
\toprule
& & \multicolumn{5}{c}{NRMSE} & Mean fitting\\
& & Noise-free & \multicolumn{2}{c}{1\% noise} & \multicolumn{2}{c}{5\% noise} & time (min)\\
\cmidrule(lr){4-5}\cmidrule(lr){6-7}
Representation & Fitting method & Raw & Raw & Smoothed & Raw & Smoothed & \\
\midrule
\multicolumn{8}{l}{\textit{Two-mass system with constant damping and unknown input map}}\\
Analytic & Gradient-based/scheme & 4.02e-2 & 4.61e-2 & 4.26e-2 & 5.76e-2 & 4.86e-2 & 153\\
Analytic & Gradient-based/roll-out & 4.36e-2 & 4.44e-2 & 4.49e-2 & 7.31e-2 & 5.33e-2 & 463\\
Analytic & Regression/derivative & 4.13e-2 & 4.34e-2 & 3.97e-2 & 2.63e-1 & 6.11e-2 & 3.01\\
Analytic & Regression/midpoint & 3.94e-2 & 7.44e-2 & 4.00e-2 & 5.07e-1 & 6.05e-2 & 3.15\\
Neural network & Gradient-based/scheme & 1.10e-2 & 5.15e-2 & 2.89e-2 & 1.46e-1 & 7.92e-2 & 29.9\\
Neural network & Gradient-based/roll-out & 1.72e-2 & 2.98e-2 & 2.61e-2 & 7.97e-2 & 5.87e-2 & 89.9\\
Hybrid B & Gradient-based/scheme & 4.27e-2 & 7.43e-2 & 5.18e-2 & 1.54e-1 & 8.78e-2 & 54.6\\
Hybrid B & Gradient-based/roll-out & 4.20e-2 & 5.38e-2 & 4.35e-2 & 9.05e-2 & 8.30e-2 & 164\\
Hybrid C & Gradient-based/scheme & \textbf{6.01e-5} & 1.18e-2 & \textbf{3.28e-3} & 3.48e-2 & \textbf{1.93e-2} & 54.5\\
Hybrid C & Gradient-based/roll-out & 1.59e-4 & 1.10e-2 & 6.40e-3 & 5.34e-2 & \textbf{1.78e-2} & 164\\
\midrule
\multicolumn{8}{l}{\textit{Two-mass system with state-dependent damping and known input map}}\\
Analytic & Gradient-based/scheme & 1.51e-3 & 1.05e-2 & 5.75e-3 & 3.17e-2 & 2.67e-2 & 153\\
Analytic & Gradient-based/roll-out & 2.76e-3 & 7.77e-3 & 6.72e-3 & 5.56e-2 & 2.67e-2 & 463\\
Analytic & Regression/derivative & 9.09e-3 & 3.67e-2 & 7.46e-3 & 2.98e-1 & 7.27e-2 & 32.8\\
Analytic & Regression/midpoint & 3.86e-3 & 1.04e-1 & 8.58e-3 & 6.35e-1 & 7.59e-2 & 31.2\\
Neural network & Gradient-based/scheme & 8.45e-2 & 1.82e-1 & 1.62e-1 & 2.19e-1 & 1.80e-1 & 27.2\\
Neural network & Gradient-based/roll-out & 6.14e-2 & 1.24e-1 & 1.51e-1 & 1.55e-1 & 1.79e-1 & 87.6\\
Hybrid A & Gradient-based/scheme & \textbf{3.88e-4} & 1.10e-2 & \textbf{5.02e-3} & 3.67e-2 & 2.64e-2 & 54.4\\
Hybrid A & Gradient-based/roll-out & 8.34e-4 & 7.60e-3 & 6.91e-3 & 3.96e-2 & \textbf{2.13e-2} & 165\\
\midrule
\multicolumn{8}{l}{\textit{Nonlinear electrical circuit with constant damping and unknown input map}}\\
Analytic & Gradient-based/scheme & 4.01e-1 & 3.91e-1 & 3.96e-1 & 3.86e-1 & 4.03e-1 & 153\\
Analytic & Gradient-based/roll-out & 3.86e-1 & 3.85e-1 & 3.87e-1 & \textit{failed}\textsuperscript{*} & 4.03e-1 & 463\\
Analytic & Regression/derivative & \textit{failed}\textsuperscript{\textdagger} & \textit{failed}\textsuperscript{\textdagger} & \textit{failed}\textsuperscript{\textdagger} & \textit{failed}\textsuperscript{\textdagger} & \textit{failed}\textsuperscript{\textdagger} & 1.80\\
Analytic & Regression/midpoint & \textit{failed}\textsuperscript{\textdagger} & \textit{failed}\textsuperscript{\textdagger} & \textit{failed}\textsuperscript{\textdagger} & \textit{failed}\textsuperscript{\textdagger} & \textit{failed}\textsuperscript{\textdagger} & 1.77\\
Neural network & Gradient-based/scheme & \textbf{4.13e-3} & 5.73e-2 & 2.77e-2 & 9.95e-2 & 8.48e-2 & 29.9\\
Neural network & Gradient-based/roll-out & 5.05e-3 & 3.38e-2 & \textbf{2.13e-2} & 9.06e-2 & \textbf{6.77e-2} & 89.9\\
Hybrid B & Gradient-based/scheme & 6.06e-2 & 8.43e-2 & 6.84e-2 & 1.12e-1 & 1.07e-1 & 54.5\\
Hybrid B & Gradient-based/roll-out & 5.78e-2 & 7.45e-2 & 7.29e-2 & 1.27e-1 & 1.23e-1 & 165\\
Hybrid C & Gradient-based/scheme & 3.80e-1 & 3.72e-1 & 3.82e-1 & 3.59e-1 & 3.79e-1 & 54.5\\
Hybrid C & Gradient-based/roll-out & 4.39e-1 & 4.42e-1 & 4.41e-1 & 4.41e-1 & 4.39e-1 & 165\\
\bottomrule
\end{tabular}
\par\smallskip
\begin{minipage}{\textwidth}\footnotesize
The errors are the average over the three datasets. The fitting times are the average over all five conditions and three datasets, including unsuccessful fits.
\textsuperscript{*}At least one test forecast failed.
\textsuperscript{\textdagger}At least one fit failed validation.
\end{minipage}
\end{table*}

\subsection{Impact of the choice of function candidates}
\label{sec:libraries}
The performance of the analytic models is naturally very sensitive to the choice of the candidate functions in the library. Table~\ref{tab:controls} presents results using alternative libraries to those considered in Table~\ref{tab:three-noise}. For the state-dependent damping in the two-mass system, the alternative library contains a constant, \(p_i^2\), and three Gaussian functions \(\exp[-p_i^2/s^2]\), with width scales \(s\in\{0.25,0.5,1\}\). Note that the library thus contains the true damping law. For the electrical circuit example, we add Gaussian functions of $q$ and $\phi$ to the polynomial Hamiltonian library, and Gaussian functions of $q$ to the input-gain library. Inspired by \citet{najnudel}, we choose the centres and widths based on the observed training range. For each coordinate, we use seven equally spaced centres between the 1st and 99th percentiles of its raw training values.

\begin{figure}[tb]
\centering
\includegraphics{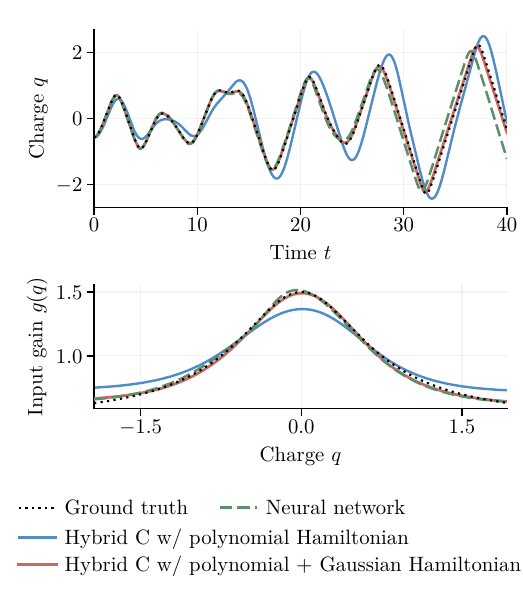}
\caption{Ground truth and model predictions for the first test trajectory and input gain of the first dataset of the circuit example. All models were fitted to smoothed data with 5\% noise using the integration-scheme loss.}
\label{fig:circuit-trajectory}
\end{figure}

The results in Table~\ref{tab:controls} for the two-mass system with state-dependent damping confirm that on noise-free data, including the ground truth in the candidate-function library gives a low error. However, on data with 5\% added noise, the hybrid model with a neural network modelling the damping still outperforms the pure analytic model. For the electrical circuit example, we see that adding Gaussian functions to the candidate-function libraries gives analytic and hybrid models that outperform the neural-network models. Figure~\ref{fig:circuit-trajectory} demonstrates how extending hybrid C's Hamiltonian library with Gaussian functions improves both the charge prediction and the learned input gain, even though its input-gain representation is unchanged.

\begin{table}[t]
\centering
\caption{Prediction error with alternative candidate-function libraries.}
\label{tab:controls}
\setlength{\tabcolsep}{3pt}
\begin{tabular}{lrrr}
\toprule
& \multicolumn{3}{c}{Prediction error}\\
Fitting method & Noise-free & 1\% noise & 5\% noise\\
\midrule
\multicolumn{4}{l}{\textit{Two-mass system with state-dependent damping}}\\
\multicolumn{4}{l}{Analytic model w/ true damping widths and centres}\\
Gradient-based/scheme & 1.04e-6 & \textbf{4.66e-3} & 2.70e-2\\
Gradient-based/roll-out & \textbf{2.56e-7} & \textbf{4.54e-3} & 2.56e-2\\
Regression/derivative & 8.56e-3 & 5.73e-3 & 8.00e-2\\
Regression/midpoint & 3.95e-3 & 6.12e-3 & 8.05e-2\\
\midrule
\multicolumn{4}{l}{\textit{Nonlinear electrical circuit}}\\
\multicolumn{4}{l}{Analytic w/ Gaussian Hamiltonian and gain extensions}\\
Gradient-based/scheme & 7.39e-3 & 2.30e-2 & 6.04e-2\\
Gradient-based/roll-out & 1.50e-2 & 2.08e-2 & 6.05e-2\\
Regression/derivative & 5.33e-3 & \textbf{1.32e-2} & 7.31e-2\\
Regression/midpoint & \textbf{3.73e-3} & \textbf{1.30e-2} & 8.43e-2\\
\addlinespace
\multicolumn{4}{l}{Hybrid B w/ Gaussian gain extension}\\
Gradient-based/scheme & 5.89e-3 & 2.60e-2 & 9.53e-2\\
Gradient-based/roll-out & 8.40e-3 & 2.87e-2 & 7.79e-2\\
\addlinespace
\multicolumn{4}{l}{Hybrid C w/ polynomial + Gaussian Hamiltonian}\\
Gradient-based/scheme & 6.58e-3 & 1.80e-2 & \textbf{4.76e-2}\\
Gradient-based/roll-out & 9.12e-3 & 1.64e-2 & \textbf{4.44e-2}\\
\bottomrule
\end{tabular}
\par\smallskip
\begin{minipage}{\columnwidth}\footnotesize
The 1\% and 5\% noise results are on smoothed training data.
\end{minipage}
\end{table}

\subsection{Computational time}
We decided in advance how much computational time to allocate to each fitting method, as described at the end of Section \ref{sec:examples}, according to the assumed requirements of the specific method. Hence, the fitting times reported in Table~\ref{tab:three-noise} do not represent time to convergence. For analytic models, we allocated less time to regression because linear least-squares fitting generally requires less computation than nonlinear optimisation with Adam. 

\subsection{Energy bounds and extrapolation}
\label{sec:properties}
As stated in Section~2.1, passivity of the system relies on a lower bound on the Hamiltonian. However, in our study, we impose no global lower bound on the analytic models for the Hamiltonian. The test trajectories extend to \(t=40\), but several models with polynomial Hamiltonians fitted by gradient-based optimisation failed between \(t=20\) and \(t=40\). We therefore report prediction errors up to
\(t=20\) in the tables. The neural-network Hamiltonian models are bounded above and below for finite weights because their hidden units use bounded activation functions, and the neural-network and hybrid B models are thus guaranteed to be passive. Passivity alone does not guarantee bounded trajectories or accurate extrapolation, but stronger constraints on the Hamiltonian can guarantee stability and bounded unforced trajectories \citep{roth}.

\section{Conclusion}
Our study shows that hybrid models that combine candidate-function and neural-network components can be beneficial over pure analytic models or pure neural-network models, depending on the underlying system and the choice of candidate-function library. Other findings include: smoothing the training data generally improves prediction accuracy, and roll-out fitting generally gives similar or moderately lower prediction errors than scheme fitting, despite having three times the fitting budget.

\section*{Declaration of generative AI and AI-assisted technologies in the writing process}
ChatGPT/Codex (OpenAI) and Claude (Anthropic) were used to assist with setting up experiments and drafting the manuscript. The authors reviewed all AI-assisted contributions and take full responsibility for the publication.

\bibliography{references}

@article{eidnes,
  author = {Eidnes, S. and Stasik, A. J. and Sterud, C. and B{\o}hn, E. and Riemer-S{\o}rensen, S.},
  title = {Pseudo-{Hamiltonian} neural networks with state-dependent external forces},
  journal = {Physica D},
  volume = {446},
  pages = {133673},
  year = {2023},
}

@article{holmsen,
  author = {Holmsen, S. and Eidnes, S. and Riemer-S{\o}rensen, S.},
  title = {Pseudo-{Hamiltonian} system identification},
  journal = {J. Comput. Dyn.},
  volume = {11},
  number = {1},
  pages = {59--91},
  year = {2024},
}

@misc{myhr,
  author = {Myhr, H. N. and Eidnes, S. and Kutz, J. N.},
  title = {{SIPHy}: Sparse identification of port-{Hamiltonian} systems from noisy data},
  year = {2026},
  howpublished = {arXiv preprint, arXiv:2609.20399v2}
}

@misc{meltzer,
  author = {Meltzer, S. and Eidnes, S. and Stasik, A. J.},
  title = {Learning partially observed systems with neural {Hamiltonian} ordinary differential equations},
  year = {2026},
  howpublished = {arXiv preprint, arXiv:2605.23510v1}
}

@article{moradi,
  author = {Moradi, S. and Beintema, G. I. and Jaensson, N. O. and T{\'o}th, R. and Schoukens, M.},
  title = {Port-{Hamiltonian} neural networks with output error noise models},
  journal = {Automatica},
  volume = {187},
  pages = {112892},
  year = {2026},
}

@inproceedings{lee,
  author = {Lee, K. and Trask, N. and Stinis, P.},
  title = {Structure-preserving sparse identification of nonlinear dynamics for data-driven modeling},
  booktitle = {Proceedings of Mathematical and Scientific Machine Learning},
  editor = {Dong, B. and Li, Q. and Wang, L. and Xu, Z.-Q. J.},
  series = {Proceedings of Machine Learning Research},
  publisher = {PMLR},
  volume = {190},
  pages = {65--80},
  year = {2022},
}

@inproceedings{greydanus,
  author = {Greydanus, S. and Dzamba, M. and Yosinski, J.},
  title = {{Hamiltonian} neural networks},
  booktitle = {Advances in Neural Information Processing Systems},
  volume = {32},
  year = {2019},
}

@article{desai,
  author = {Desai, S. A. and Mattheakis, M. and Sondak, D. and Protopapas, P. and Roberts, S. J.},
  title = {Port-{Hamiltonian} neural networks for learning explicit time-dependent dynamical systems},
  journal = {Phys. Rev. E},
  volume = {104},
  number = {3},
  pages = {034312},
  year = {2021},
}

@article{duong,
  author = {Duong, T. and Altawaitan, A. and Stanley, J. and Atanasov, N.},
  title = {Port-{Hamiltonian} neural {ODE} networks on {Lie} groups for robot dynamics learning and control},
  journal = {IEEE Trans. Robot.},
  volume = {40},
  pages = {3695--3715},
  year = {2024},
}

@article{brunton,
  author = {Brunton, S. L. and Proctor, J. L. and Kutz, J. N.},
  title = {Discovering governing equations from data by sparse identification of nonlinear dynamical systems},
  journal = {Proc. Natl. Acad. Sci. USA},
  volume = {113},
  number = {15},
  pages = {3932--3937},
  year = {2016},
}

@article{castanos,
  author = {Casta{\~n}os, F. and Jayawardhana, B. and Ortega, R. and Garc{\'i}a-Canseco, E.},
  title = {Proportional plus integral control for set-point regulation of a class of nonlinear {RLC} circuits},
  journal = {Circuits, Systems, and Signal Processing},
  volume = {28},
  number = {4},
  pages = {609--623},
  year = {2009},
}

@article{cherifi,
  author = {Cherifi, K. and El Messaoudi, A. and Gernandt, H. and Roschkowski, M.},
  title = {Nonlinear port-{Hamiltonian} system identification from input-state-output data ({ISO-pHNN})},
  journal = {Physica D},
  volume = {497},
  pages = {135368},
  year = {2026},
}

@inproceedings{roth,
  author = {Roth, F. J. and Klein, D. K. and Kannapinn, M. and Peters, J. and Weeger, O.},
  title = {Stable port-{Hamiltonian} neural networks},
  booktitle = {Advances in Neural Information Processing Systems},
  volume = {38},
  year = {2025},
}

@inproceedings{khosrovian,
  author = {Khosrovian, R. A. and Yaguchi, T. and Matsubara, T.},
  title = {Port-{Hamiltonian} neural networks for learning coupled systems and their interactions},
  booktitle = {Machine Learning and the Physical Sciences Workshop at NeurIPS},
  year = {2024},
}

@article{vanderschaft,
  author = {van der Schaft, A. and Jeltsema, D.},
  title = {Port-{Hamiltonian} systems theory: An introductory overview},
  journal = {Found. Trends Syst. Control},
  volume = {1},
  number = {2--3},
  pages = {173--378},
  year = {2014},
}

@article{celledoni,
  author = {Celledoni, E. and Eidnes, S. and Myhr, H. N.},
  title = {Learning dynamical systems from noisy data with inverse-explicit integrators},
  journal = {Physica D},
  volume = {472},
  pages = {134471},
  year = {2025},
}

@inproceedings{najnudel,
  author = {Najnudel, J. and M{\"u}ller, R. and H{\'e}lie, T. and Roze, D.},
  title = {Identification of nonlinear circuits as port-{Hamiltonian} systems},
  booktitle = {Proceedings of the 24th International Conference on Digital Audio Effects},
  pages = {1--8},
  year = {2021},
}

@inproceedings{chen,
  author = {Chen, R. T. Q. and Rubanova, Y. and Bettencourt, J. and Duvenaud, D. K.},
  title = {Neural ordinary differential equations},
  booktitle = {Advances in Neural Information Processing Systems},
  volume = {31},
  year = {2018},
}
\end{document}